\documentclass[letters,usenatbib]{mnras}

\usepackage{mathptmx}

\usepackage[T1]{fontenc}
\usepackage{ae,aecompl}

\usepackage{graphicx}	
\usepackage{amsmath}	
\usepackage{amssymb}	
\usepackage{wasysym}           
\usepackage{epstopdf}
\usepackage{mathrsfs}
\usepackage{natbib}
\usepackage{color}
\usepackage{hyperref}

\title[Circumbinary discs from tidal disruption events]{Circumbinary discs from tidal disruption events}

\author[Coughlin \& Armitage]{
Eric R. Coughlin,$^{1}$\thanks{email: eric\_coughlin@berkeley.edu}\thanks{Einstein fellow}
Philip J. Armitage,$^{2,3}$
\\
$^{1}$Astronomy Department and Theoretical Astrophysics Center, University of California, Berkeley, Berkeley, CA 94720 \\
$^{2}$JILA, University of Colorado and National Institute of Standards and Technology, 440 UCB, Boulder, CO 80309-0440, USA \\
$^{3}$Department of Astrophysical and Planetary Sciences, University of Colorado, 391 UCB, Boulder, CO 80309-0391, USA\\
}

\date{Accepted XXX. Received YYY; in original form ZZZ}

\pubyear{2016}

\begin{document}
\label{firstpage}
\pagerange{\pageref{firstpage}--\pageref{lastpage}}
\maketitle

\begin{abstract}
Tidal disruption events, which occur when a star is shredded by the tidal field of a supermassive black hole, provide a means of fueling black hole accretion. Here we show, using a combination of three body orbit integrations and hydrodynamic  simulations, that these events are also capable of generating circumbinary rings of gas around tight supermassive black hole binaries with small mass ratios. Depending on the thermodynamics, these rings can either fragment into clumps that orbit the binary, or evolve into a gaseous circumbinary disc. We argue that tidal disruptions  provide a direct means of generating circumbinary discs around supermassive black hole binaries and, more generally, can replenish the reservoir of gas on very small scales in galactic nuclei.
\end{abstract}




\section{Introduction}
Supermassive black hole (SMBH) binaries form at pc-scales from galactic collisions and subsequent dynamical friction (e.g., \citealt{milos03,kelley17}). The presence of gas surrounding these binaries may play a role in their subsequent merger \citep{begelman80,cuadra09,goicovic16a,goicovic16b}, and is a pre-requisite to attempts to identify close binaries through, for example, searches for periodic emission from Active Galactic Nuclei \citep[AGN;][]{graham15,charisi16,dorazio17}. On still smaller scales, gas, in the form of a dense accretion disc, is needed if there is any prospect to observe electromagnetic counterparts to black hole mergers identified through gravitational wave emission \citep{armitage02,rossi10,schnittman11,giacomazzo12,cerioli16}. To date, however, very few close binaries have been directly identified \citep{rodriguez06}, while constraints from pulsar timing bound only a subset of realistic models \citep{arzoumanian16,chen17}. As a result we know neither the number of close binaries, nor the fraction that host gas discs.

Sources of gas at the very centre of galactic nuclei range from the strong organized or chaotic gas flows implicated in the fueling of high-luminosity AGN \citep{shlosman89,king06,capelo15}, down to the meagre baseline set by stellar wind fueling in the Galactic Centre \citep{cuadra06}. Tidal disruption events \citep[TDEs;][]{rees88}, when a star is destroyed by the tidal field of the SMBH, represent an intermediate source. Given TDE rates of the order of $10^{-5} - 10^{-4} \ {\rm gal}^{-1} \ {\rm yr}^{-1}$ \citep{frank76,stone16}, the time-averaged accretion rate corresponds to  $L \sim 5 \times 10^{-5} - 5 \times 10^{-4} \ L_{\rm Edd}$ (for a $10^6 \ M_\odot$ black hole and $1 \ M_\odot$ accreted per TDE at a radiative efficiency of $\epsilon = 0.1$). At the high end of the rate estimate the amount of mass accreted constitutes a significant fraction of the total mass needed to assemble a low mass black hole \citep{milos06}. Gas supply by TDEs is  stochastic, with random disruptions supplying mass (to a single black hole) at an initially often super-Eddington rate that is followed by a steep decay $\dot{M} \propto t^{-5/3}$ \citep{phinney89,evans89}. The average fueling rate depends on the nuclear dynamics \citep{stone16}, but is largely independent of the galactic properties (large gas masses, bars, close-in massive stars, etc) that set other accretion processes.

In this paper we show that, in galactic nuclei containing SMBH binaries with a massive primary black hole and an extreme mass ratio, TDEs can directly form a circumbinary disc which, depending upon the thermodynamics of the gas, may or may not fragment \citep{coughlin15,coughlin16}. We suggest that stars encountering extreme mass ratio binaries provide a novel channel for forming a long-lived gas reservoir on sub-pc scales.

\section{Simulations}
\label{sec:simulations}
We consider a circular binary SMBH system with $q = M_2/M_1 = 0.005$, $M_1 = 10^8M_{\odot}$, and $a = 5$ mpc, the gravitational-wave inspiral time of which is $\tau_{\rm GW} \simeq 6.5\times10^{7}$ yr. \textcolor{black}{Such an extreme mass ratio would be uncommon for systems with total black hole mass $\lesssim 10^6 M_\odot$. For binary systems where the total mass is constrained to exceed $\sim 10^8 M_\odot$, however, results from an analysis of mergers in the Illustris simulations suggest that mergers with $q=10^{-2} - 10^{-3}$ are frequent \citep{kelley17}. \citet{rodriguez-gomez15} also found from the Illustris simulations that mergers involving larger-mass galaxies preferentially occurred with smaller-mass companions; adopting any of the usual galaxy-SMBH correlations then naturally produces a substantial number of low-$q$ SMBH binaries with relatively massive primaries.} The orbit of a star is first followed by integrating the restricted 3-body equations, starting with the test particle initialized randomly over a sphere of radius $50a$ on a parabolic orbit with pericenter uniformly distributed between $0$ and $2a$.  If the star comes within the tidal radius of the secondary, we then trace its orbit back to when it was $5r_t$ from the SMBH and initialize a full hydrodynamics simulation in {\sc phantom} \citep{price17}. The star is modeled as a polytrope, with either $\gamma = 5/3$ or $\gamma = 1.55$, by placing $5\times10^5$ particles on an appropriately-stretched, close-packed sphere, and is relaxed in isolation for ten sound-crossing times prior to disruption. We include self-gravity using a k-D tree \citep{gafton11}, and evolve the gas pressure adiabatically, with the adiabatic index set to the polytropic index. 

The effects of two different adiabatic indices are investigated because the debris stream resulting from a TDE can be gravitationally unstable depending on the equation of state of the gas \citep{coughlin15,coughlin16}. Specifically, if $\gamma \ge 5/3$, the decrease in the gas pressure results in a thin, quasi-hydrostatic stream width, the gravitational collapse timescale of which is shorter than the dynamical time. If $\gamma < 5/3$, the pressure decreases less rapidly, permitting a wider stream and stabilizing the debris against fragmentation. The adiabatic index of the gas could be less than $5/3$ if radiative recombination heats the material \citep{kasen10} or magnetic pressure is important \citep{guillochon17,bonnerot16}, while a larger $\gamma$ is possible if cooling is efficient. 

We integrated $\sim 6\times10^6$ restricted 3-body orbits, $>99\%$ of which resulted in ejected stars or stars swallowed whole by the primary. There were 44 that were disrupted by the secondary, and \citet{coughlin17b} investigated the hydrodynamics of those where prompt accretion onto the SMBH occurred. Instead, here we focus on one of the cases where the stream was not ejected entirely, but was placed on a loosely-bound orbit about the binary. The motion of the gas is followed for $\sim 6.5$ years (or just over two binary orbits) when $\gamma = 5/3$, and for $\sim 22$ years when $\gamma = 1.55$. 

\begin{figure*}
\includegraphics[width=1.00\textwidth, height=0.525\textwidth]{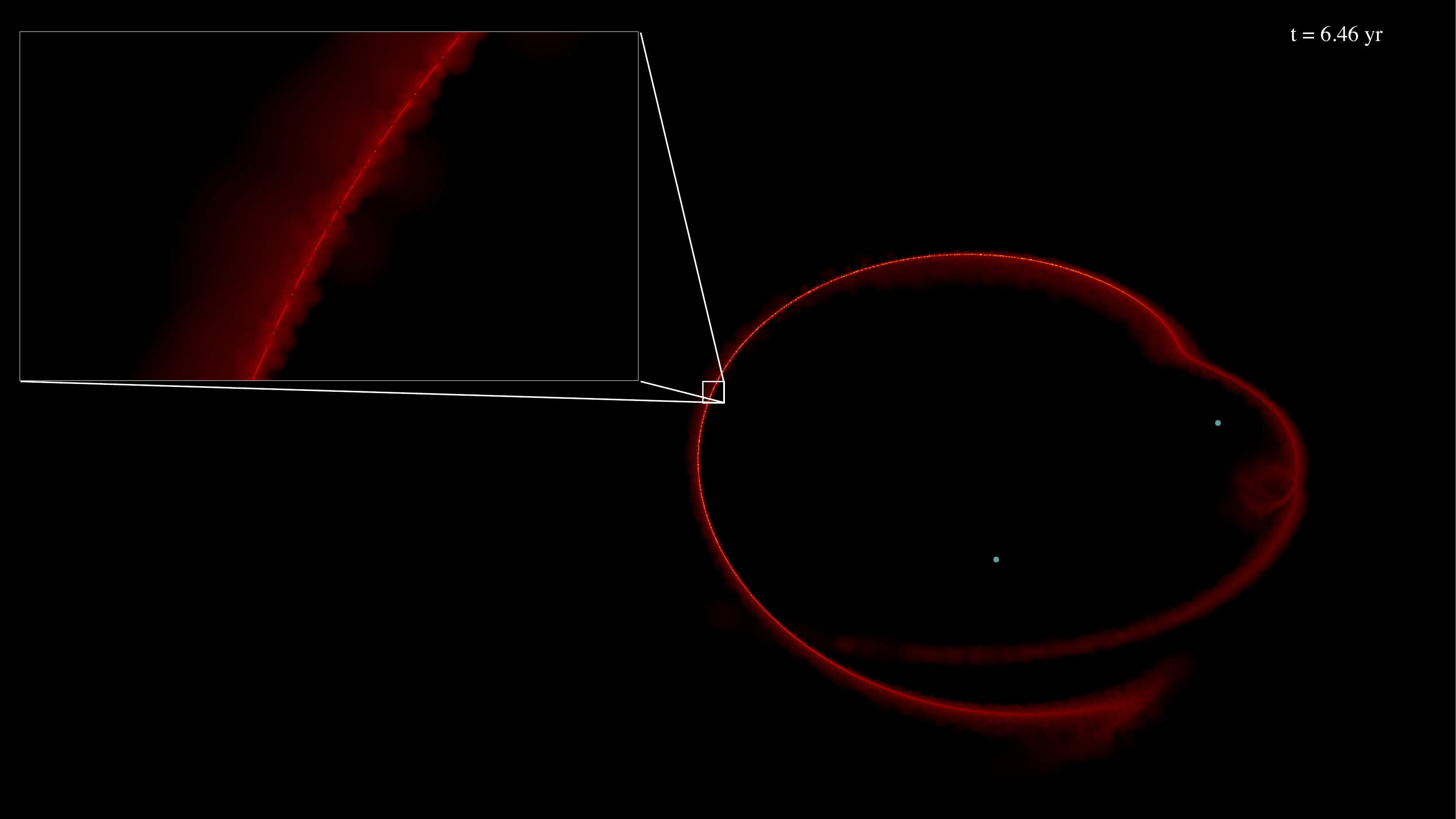}
\includegraphics[width=1.00\textwidth, height=0.525\textwidth]{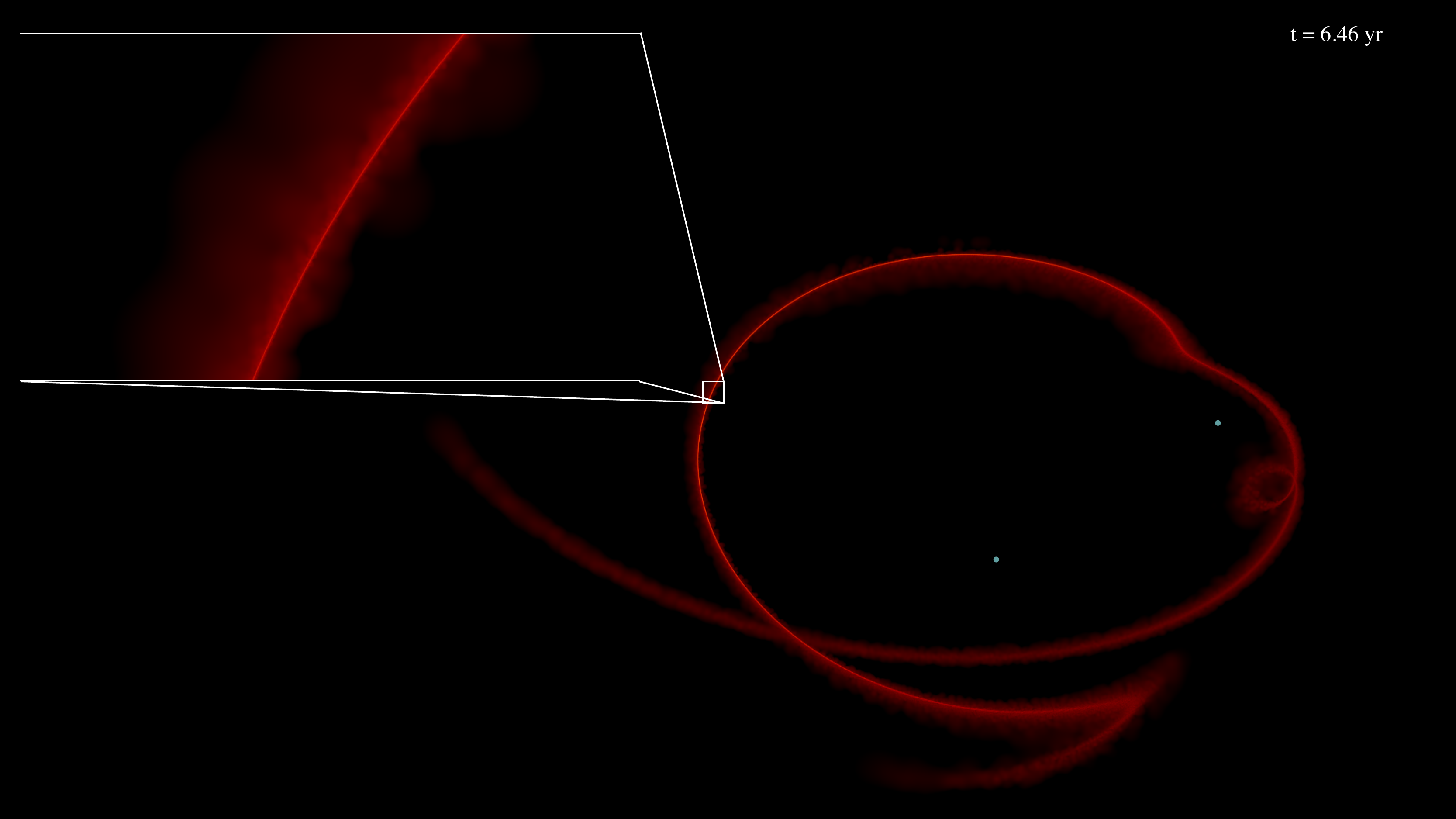}
\caption{The debris stream generated from the tidal disruption event at $\sim 6.5$ years (slightly more than two binary orbits) for $\gamma = 5/3$ -- top panel -- and $\gamma = 1.55$ -- bottom panel. Colors trace the log of the density of the gas, with brighter colors indicating denser regions. In each case the stream is wound into a ring that surrounds the binary, resulting in no immediate accretion. The insets show closeups of the stream; for $\gamma = 5/3$, the stream fragments into a number of clumps, while the $\gamma = 1.55$ case remains smooth and somewhat thicker.}
\label{fig:streams}
\end{figure*}

Figure \ref{fig:streams} illustrates the distribution of the tidally-disrupted debris from the simulation with $\gamma = 5/3$ (top panel) and $\gamma = 1.55$ (bottom panel) at a time of $\sim 6.5$ years post-disruption; colors trace the log of the density, with brighter colors highlighting denser regions. TDEs in binaries typically result  in prompt fallback toward the disrupting hole, the formation of a small-scale disc, and, over a timescale of a few binary orbits, an accretion rate that decays as a power-law with intermittent rebrightenings (see \citealt{coughlin17,coughlin17b}). In this instance, by contrast, the debris stream generated from this event does not accrete onto either black hole. Instead, the material is swung into a long arc that, after only two binary orbits, encircles both black holes in a circumbinary ring.

\begin{figure*}
\includegraphics[width=0.495\textwidth, height=0.25\textwidth]{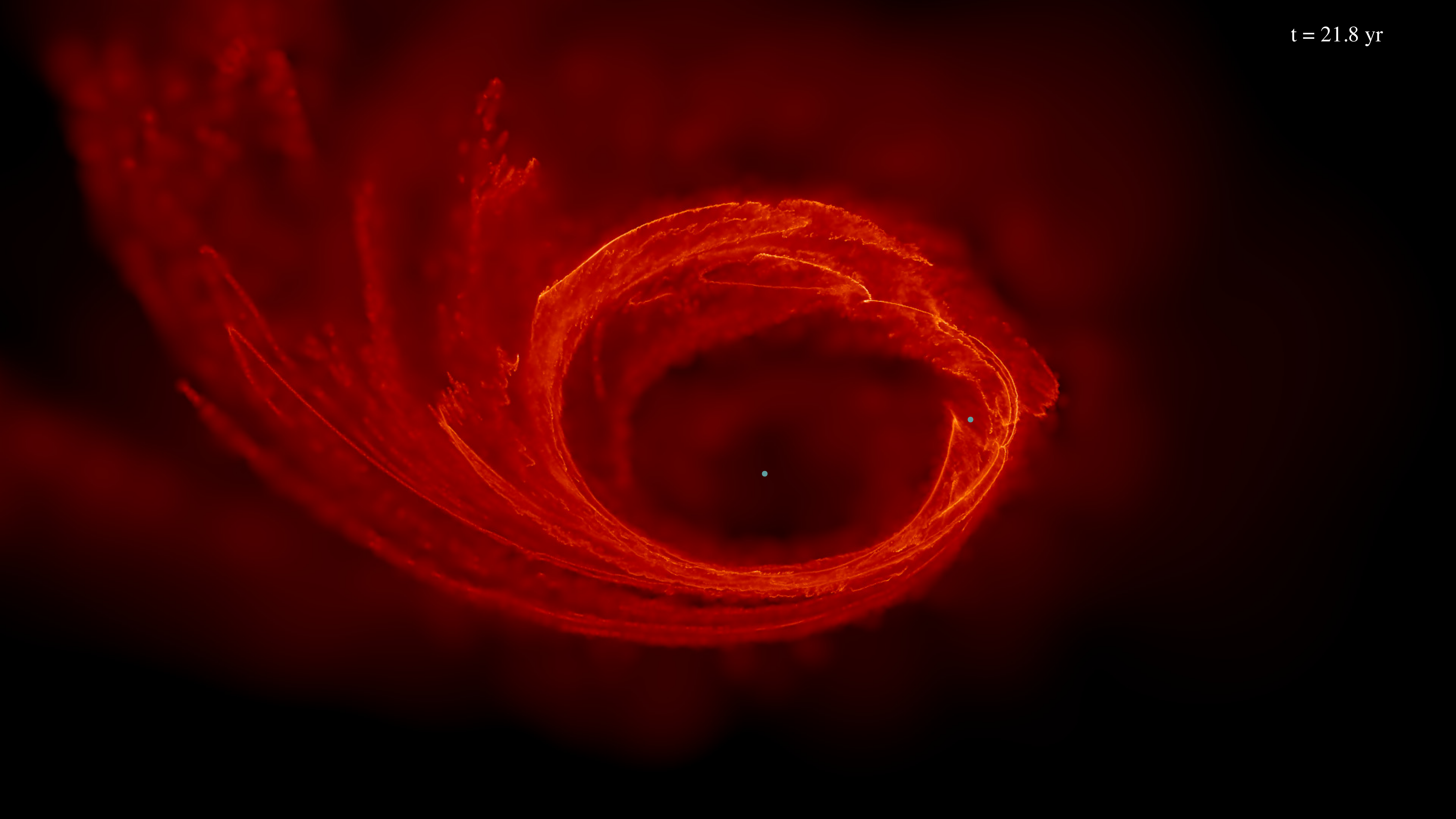}
\includegraphics[width=0.495\textwidth, height=0.25\textwidth]{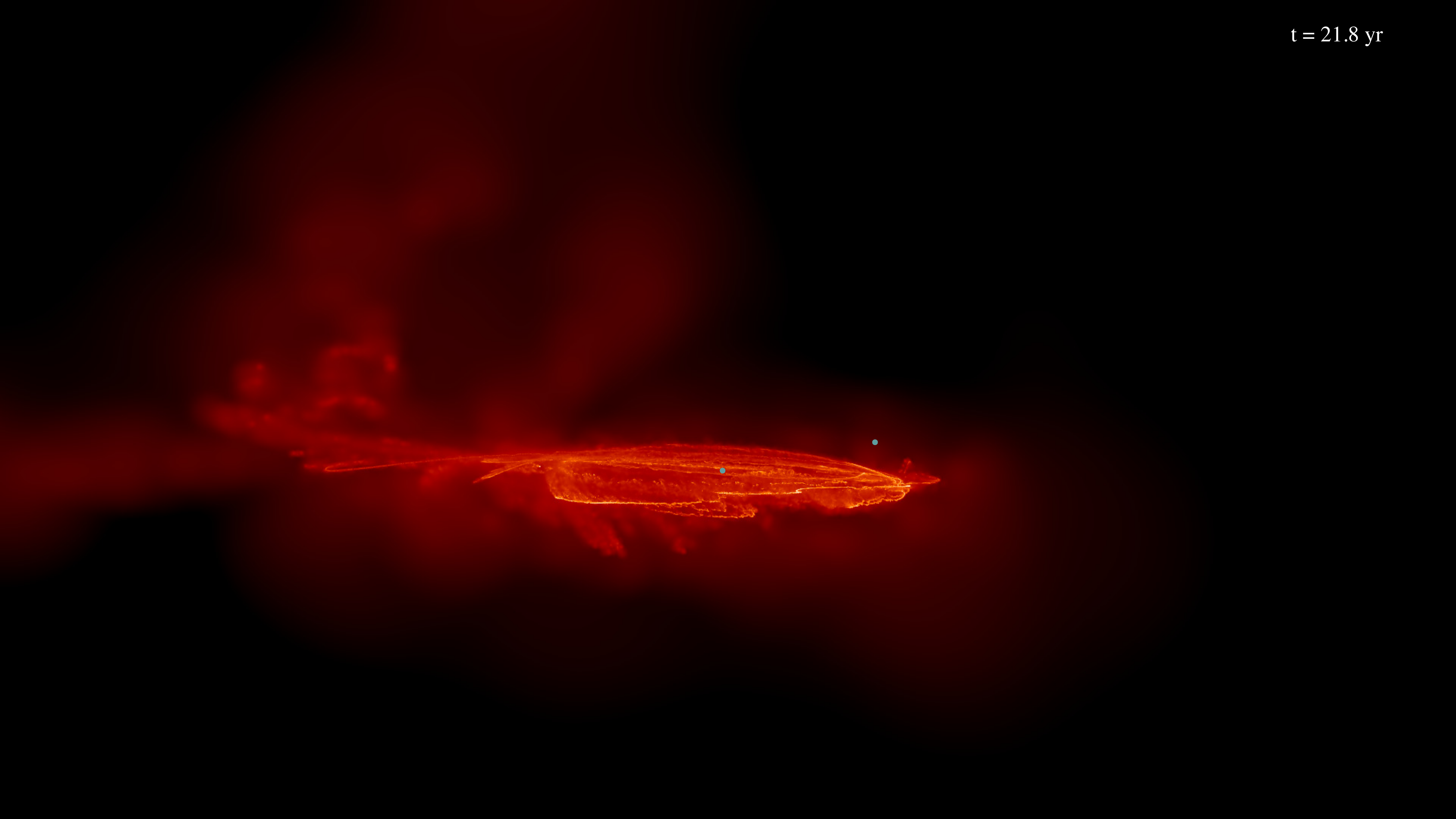}
\caption{The debris from the TDE at a time of \textcolor{black}{$\sim 22$ yr post-disruption, or nearly 7 binary orbits,} when $\gamma=1.55$; the left-hand panel shows the projection onto the binary plane, \textcolor{black}{ and the middle panel is inclined by $45^{\circ}$ to the binary plane}. By this time the configuration of the gas has become more chaotic due to perturbations induced by the secondary, \textcolor{black}{and the material has spread into a thin disc that, as indicated by the middle panel, has an inclination of roughly 45$^{\circ}$ to the binary orbital plane.}}
\label{fig:disk}
\end{figure*}

The inset in these figures shows a closeup view of a segment of the stream. As expected \citep{coughlin16}, when $\gamma = 5/3$, the stream becomes punctuated by a series of small-scale clumps, which are regions where the material has collapsed under its own self-gravity. For $\gamma = 1.55$, on the other hand, the inset demonstrates that the stream remains wider, and the pressure support provides stability against the fragmentation that occurs for larger adiabatic indices. 

Figure \ref{fig:disk} shows the distribution of the material after \textcolor{black}{$\sim 22$ years, or roughly seven orbits, after the initial disruption, with the left-hand panel projected onto the binary orbital plane and the middle panel inclined by $45^{\circ}$.} By this point the stream has expand laterally into a disc of debris. The lateral expansion is due to perturbations caused by the binary and the viscosity employed by the numerical method. There is still, however, very little accretion at this time.

\textcolor{black}{\citet{coughlin17} analyzed the hydrodynamic evolution of 120 disruptions by a $q = 0.2$ SMBH binary, and in none of their simulations did they find a circumbinary disc. We posit that the small-q binary considered here is more likely to form a disc because most disrupted stars first enter weakly-bound orbits about the primary, and then gradually diffuse in energy space due to perturbations from the motion of the primary and secondary. Thus, by the time a star is disrupted by a small-q binary, its specific energy and angular momentum can be comparable to those of the binary itself, enabling disrupted material to be placed on approximately circular orbits that contrast the usual, nearly-parabolic ones obtained from a standard TDE. }

\section{Long term evolution}
\label{sec:implications}
The fate of the gas generated from the tidal disruption event is dependent on its equation of state. When the effective adiabatic index $\gamma$ satisfies $\gamma \ge 5/3$, the debris collapses under its own self-gravity to form isolated pockets, or clumps, of material, while softer equations of state retain a smooth structure. We consider the implications of each of these cases.

\subsection{Efficient cooling and fragmentation}
In the absence of any pre-existing gas reservoir, the clumps generated from the circumbinary ring would form a distinct population of objects bound to the binary. Since the unstable modes grow as power-laws instead of exponentials when $\gamma = 5/3$ \citep{coughlin16b}, the instability is  difficult to resolve numerically, and the properties of the clumps (e.g., mass and time of formation) depend on the resolution of the simulation \citep{coughlin15}. It is therefore uncertain as to what these clumps could condense into over longer timescales (e.g., brown dwarfs or  planetary mass objects).

Nevertheless, these clumps will likely all form around the same time with similar physical and dynamical properties, and their subsequent evolution is not strongly dependent on their mass. The circular restricted three body problem is a good approximation to the dynamics for time scales on which shrinkage of the binary due to gravitational waves is negligible. In the test particle limit, there is a precise boundary between orbits that are Hill stable (and hence guaranteed to never encounter either of the black holes, though escape remains possible) and those that are not. Most (but not all) of the orbits that fail to satisfy the criterion for Hill stability are unstable, in the sense that they eventually lead to close encounters with either the primary or the secondary. In the test particle limit, the long term outcome of circumbinary disc fragmentation is thus expected to be a burst of disruptions of low mass objects (as the unstable population is cleared out), leaving behind a population of Hill stable objects, some of which may ultimately be ejected.

Relaxation can in principle modify the dynamics, but this effect is negligible for the low masses of interest here. For a disc of objects with total mass $M_{\rm tot}$ and individual mass $m$, orbiting at radius $r$ in an annulus of extent $\Delta r = \epsilon r$ with velocity dispersion $\sigma$, the 2-body relaxation time is estimated to be \citep[e.g.][]{alexander07},
\begin{equation}
 t_{\rm relax} = \frac{C_1 \epsilon r^2 \sigma^4}{G^2 M_{\rm tot}m \ln \Lambda} t_{\rm orb}.
\end{equation} 
Here $t_{\rm orb}$ is the orbital time, $C_1$ is an  order unity constant, and $\ln \Lambda$ is the Coulomb logarithm. Noting that $\sigma \approx e v_K$, where $e$ is the eccentricity and $v_K$ the orbital velocity, we obtain for a ring ($\epsilon = 0.1$, taking $C_1 = 2$ and $\ln \Lambda = 20$),
\begin{equation}
 t_{\rm relax} \sim 10^{-2} \left( \frac{M_1^2}{M_{\rm tot}m} \right) e^4 t_{\rm orb}.
\end{equation}
This is very rough, but for reasonable parameters the conclusion is that 2-body relaxation for the very low mass discs that might be produced from TDEs is securely negligible.

If there is a pre-existing disc surrounding the binary, the Kelvin-Helmholtz instability could prevent the formation of the clumps, and the debris stream from the TDE may simply dissolve into and add mass to the background disc \citep{bonnerot16b}. However, if they do have time to fragment out of the stream, their densities would likely be orders of magnitude above that of the ambient disc. Depending on the relative inclination between the stream and the pre-existing disc, the clumps could conceivably carve out regions of the disc before being dynamically dragged inward.

\subsection{Inefficient cooling and disc formation}
If the gas pressure does not decline too rapidly as the stream expands, then the material remains stable against fragmentation and, as depicted in Figure \ref{fig:disk}, spreads viscously to form a thin disc. On a timescale longer than we can directly simulate, the circumbinary disc will evolve toward alignment or anti-alignment with the binary orbital angular momentum (though the large misalignment angle of $\sim 45^{\circ}$ found here could lead to disc breaking; \citealt{nixon13}). In principle such a disc will evolve viscously \citep{pringle91}, but the viscous time scale {\em if the gas remains cold} is prohibitively long. At 5~mpc (about 500 Schwarzschild radii for a $10^8 \ M_\odot$ black hole) a Solar mass of gas, distributed over an annulus with $\Delta r \sim r$, creates a disc with a surface density $\Sigma \sim 1 \ {\rm g \ cm}^{-2}$. For low temperature opacities the vertical optical depth will be of the order of unity, so we can estimate the temperature due to viscous heating assuming that the mid-plane and effective temperatures are the same. Adopting an $\alpha$ prescription the viscosity is $\nu = \alpha c_s^2 / \Omega$, where $c_s$ is the sound speed and $\Omega$ the angular velocity. The temperature then follows by equating viscous heating, $Q_+ = (9/4) \nu \Sigma \Omega^2$, to radiative cooling $Q_- = 2 \sigma T^4$. For our fiducial parameters this yields a temperature of only tens of K, and an extremely long viscous time $r^2 / \nu$. A cold circumbinary disc formed from a TDE would evolve dynamically into an equilibrium state, but would not thereafter significantly accrete.

A more likely scenario, however, is that such a small mass of gas would form a geometrically thick disc with a temperature close to the virial temperature \citep{yuan14}. (Although, the processes that lead to cold gas evaporating into a thick disc remain quite uncertain.) In this case $c_s \sim v_K$, and the viscous time scale $t_\nu \sim (\alpha \Omega)^{-1}$. A hot disc would lead to accretion on to the black holes at a rate $\dot{M} \sim 10^{-2} \ M_\odot \ {\rm yr}^{-1}$. The secondary would capture the bulk of the accretion \citep{bate97,farris14}, potentially powering a low-luminosity (given the low secondary mass) but prolonged phase of near Eddington-rate accretion.

\section{Summary and Conclusions}
\label{sec:summary}
Using hydrodynamic simulations, we have shown that the tidal disruption of stars by supermassive black hole binaries -- particularly those with small separations ($\sim 5$ mpc) and small mass ratios ($q \sim 0.005$) -- can generate  circumbinary rings of gas. Depending on the thermodynamics, the ring can either be gravitationally unstable and fragment to form a collection of clumps that orbit the binary, or can be stable and spread over time to form a thicker, circumbinary disc. 

\textcolor{black}{The disc likely does not contribute dramatically to the inspiral of such small-q SMBH binaries -- the energy carried away by ejected stars is far more significant, both in the magnitude the energy of an individual ejected star and in the relative probability of ejection versus tidal disruption.
However, the disc could provide an important reservoir of gas to accompany the final stages of the gravitational wave-induced inspiral, thereby generating an electromagnetic transient to accompany the gravitational wave emission. Since the gravitational wave strain decreases nearly-monotonically with the binary mass ratio, it is unlikely that current pulsar timing arrays will be able to single out individual systems with mass ratios as small as those considered here \citep{sesana09,schutz16}. Nevertheless, future gravitational wave observatories such as the square-kilometer array \citep{smits09} and the laser interferometer space array \citep{amaro-seoane17} could have adequate sensitivity, and a concomitant electromagnetic signal would provide near-indisputable validation of the detections. }

\textcolor{black}{If the disc remains thin, its lifetime is very long relative to the fiducial rate of $\sim 1$ TDE per $10^{4-5}$ yr. Thus, one would expect multiple TDEs to occur over the lifetime of a single disc, each generating a new disc with a different orientation (though likely somewhat aligned to the angular momentum vector of the binary, given that disrupted stars are preferentially confined to the plane of the binary; \citealt{coughlin17}). These discs could then interact, resulting in a cancellation of angular momentum and periodic accretion bursts. On the other hand, if the disc becomes geometrically thick, then the larger accretion rate implies that the disc is relatively short-lived with respect to the time between TDEs.}

In this letter we analyzed one TDE that generated a circumbinary gas reservoir, and we did not assess the frequency with which this outcome -- as opposed to a total ejection or prompt accretion  \citep{coughlin17,coughlin17b} -- occurs. In general, we expect circumbinary flows to be formed more easily when the mass ratio is small, because the stellar specific energy and angular momentum deviate more significantly from their original values in these instances. When these deviations occur, more of the debris can remain bound to the system and the stream can orbit the binary at approximately the binary orbital frequency -- two unusual features found in the TDE investigated here. We also expect circumbinary discs formed from TDEs to be predominantly coplanar, as disruptions occur more frequently when the incoming stars are close to the orbital plane.

\section*{Acknowledgments}
Support for this work was provided by NASA through the Einstein Fellowship Program, grant PF6-170150. PJA acknowledges support from NASA through grant NNX16AI40G, and thanks the IIB at the University of Liverpool for hospitality. \textcolor{black}{We thank the referee for useful comments and suggestions. This research used the Savio computational cluster resource provided by the Berkeley Research Computing program at the University of California, Berkeley (supported by the UC Berkeley Chancellor, Vice Chancellor for Research, and Chief Information Officer). We used {\sc splash} \citep{price07} for the visualization.}

\bibliographystyle{mnras}
\bibliography{refs}

\bsp	
\label{lastpage}
\end{document}